\documentclass[
 amsmath,amssymb,
 aps,
 prl,
 twocolumn,
 superscriptaddress,
 showpacs
]{revtex4-2}

\usepackage{graphicx}
\usepackage{dcolumn}
\usepackage{bm}
\usepackage{mathtools}
\usepackage{tcolorbox}
\usepackage{pgfplots}
\usepgfplotslibrary{groupplots}
\usepackage{svg}
\usepackage{tikz}
\usepackage{makecell}
\usepackage{multirow}
\usepackage{xcolor}
\usepackage{soul}
\usepackage[colorlinks=true, allcolors=apsblue]{hyperref}
\usepackage[normalem]{ulem}

\pgfplotsset{every tick label/.append style={font=\tiny}}

\definecolor{apsblue}{RGB}{16, 38, 148}
\definecolor{mygreen}{rgb}{0.05, 0.5, 0.06}
\definecolor{mypink}{rgb}{0.96, 0.56, 0.92}
\definecolor{mybrown}{rgb}{0.34, 0.22, 0.22}
\definecolor{mypurple}{rgb}{0.45, 0, 0.81}
\definecolor{mygray}{rgb}{0.5, 0.5, 0.5}

\newcommand\br{ {\bf r}}

\newcommand\bfeta{ {\boldsymbol{\eta}}}

\newcommand\bJ{{\bf J}}

\newcommand\bk{{\bf k}}

\newcommand\bLambda{{\boldsymbol{\Lambda}}}
\renewcommand\phi\varphi 
\renewcommand\epsilon\varepsilon

\begin{document}

\preprint{APS/123-QED}

\title{Generic long-range correlations in conserved nonequilibrium mixtures}

\author{Jessica Metzger}
\affiliation{Department of Physics, Massachusetts Institute of Technology, Cambridge, Massachusetts 02139, USA}
\author{Yariv Kafri}
\affiliation{Department of Physics, Technion-Israel Institute of Technology, Haifa 3200003, Israel}
\author{Mehran Kardar}
\affiliation{Department of Physics, Massachusetts Institute of Technology, Cambridge, Massachusetts 02139, USA}
\author{Julien Tailleur}
\affiliation{Department of Physics, Massachusetts Institute of Technology, Cambridge, Massachusetts 02139, USA}

\date{\today}

\begin{abstract}
    We study correlation functions in  conserved nonequilibrium mixtures, including multi-temperature systems and non-reciprocal field theories. The corresponding linear theory is short-ranged, and nonlinearities are irrelevant in the renormalization-group sense. Nonetheless, we find that these nonlinearities generate long-ranged three-point correlations in the isotropic disordered phase. Our analytical predictions, which are based on a phenomenological theory, are confirmed by numerical simulations of Brownian colloids in contact with  thermal baths at different temperatures. Dangerously irrelevant nonlinearities in nonequilibrium mixtures thus offer a new route to long-range correlations, supporting the hypothesis that such correlations are not the exception but the rule out of equilibrium. 
\end{abstract}

\maketitle
In thermal equilibrium, short-range interactions in disordered states typically induce exponentially decaying short-range correlations. Long-range, power-law correlations occur only upon fine-tuning to a critical point. 
On the contrary, it was postulated that long-range correlations should be generic out of equilibrium~\cite{grinstein_generic_1991}. 
However, while anisotropy has been identified as generically leading to long-range correlations in conserved nonequilibrium dynamics~\cite{katz_phase_1983,katz_nonequilibrium_1984,grinstein_conservation_1990,garrido_long-range_1990,schmittmann_statistical_1995,schmittmann_driven_1998,poncet_universal_2017,ben_dor_ramifications_2019,mahdisoltani_long-range_2021,ro_disorder-induced_2021,adachi_power-law_2024}, other generic mechanisms have remained scarce.

Multi-component nonequilibrium mixtures provide an alternative setting to examine long-range correlations. 
Systems such as non-reciprocal mixtures~\cite{saha_scalar_2020,you_nonreciprocity_2020,fruchart_non-reciprocal_2021,frohoff-hulsmann_suppression_2021,frohoff-hulsmann_localized_2021,poncet_when_2022,gupta_active_2022,duan_dynamical_2023,dinelli_non-reciprocity_2023,benois_enhanced_2023,brauns_nonreciprocal_2024,pisegna_emergent_2024,greve_coexistence_2025,johnsrud_state_2025,fruchart_nonreciprocal_2026}, as well as active-passive~\cite{mccandlish_spontaneous_2012,das_phase_2016,wysocki_propagating_2016,wittkowski_nonequilibrium_2017,smrek_interfacial_2018,patteson_propagation_2018,sturmer_chemotaxis_2019,kolb_active_2020,rodriguez_phase_2020,de_castro_active_2021,de_castro_diversity_2021,williams_confinement-induced_2022,espada_burriel_active_2024,mason_dynamical_2025,venkatareddy_phase_2026} and multi-temperature fluids~\cite{awazu_segregation_2014,grosberg_nonequilibrium_2015,weber_binary_2016,smrek_small_2017,tanaka_hot_2017,ilker_phase_2020,ilker_long-time_2021,wang_tethered_2021,jardat_diffusion_2022,schwarcz_emergence_2024,mccarthy_demixing_2024,damman_algebraic_2024,venkatareddy_growth_2025} have recently attracted significant attention, primarily for their rich phase and pattern formation behavior. 
However,  less is understood about their disordered phases and the nature of correlations in homogeneous states. 
Of particular relevance is recent work that reports rapidly-decaying power-law pair correlations in a two-temperature mixture~\cite{damman_algebraic_2024}.

\begin{figure}
    \centering
    \includegraphics[]{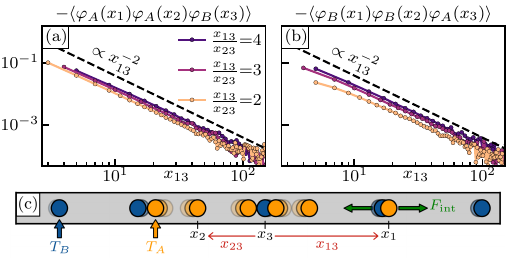}
    \caption{\textbf{Long-range 3-point correlations in a 1-dimensional 2-temperature mixture.} We simulate a mixture of Brownian particles with temperatures $T_A\neq T_B$ interacting through a soft pairwise repulsive force $F_{\rm int}(x)=-U_{\rm int}'(x)$ [panel {\bf (c)}]. {\bf (a)-(b)} Correlation between species $\alpha,\beta,\gamma$ at locations $x_1,x_2,x_3$, for (a) $\alpha,\beta,\gamma=A,A,B$ and (b) $\alpha,\beta,\gamma=B,B,A$; plotted against $x_{13}=x_1-x_3$ for different slices of $(x_{13},x_{23})$ space specified by the ratio $x_{13}/x_{23}$ (legend). Simulation parameters are $\kappa=T_A=1$, $T_B=0.1$, $N_A=N_B=200$, and $L=800$. 
    }
    \label{fig:2temp-3pt-realspace}
\end{figure}

In this Letter, we show this algebraic decay to arise from dangerously irrelevant operators. 
Crucially, these lead to genuinely long-range higher-order correlations, hence revealing a new mechanism for long-range correlations in nonequilibrium systems.
To show this, we consider the  dynamics of  \textit{conserved}, \textit{isotropic} fields $\phi_\alpha(\br,t)$ in $d$ dimensions, where $\alpha$ is the field index. Conservation implies dynamics of the form $\dot{\phi}_\alpha = -\nabla \cdot \bJ_\alpha$. The most general \textit{local} current $\bJ_\alpha$ appears as an expansion in gradients and fields as
\begin{align}
\dot{\phi}_\alpha &= \!\nabla \cdot \!\Big[ \sum_\beta \Gamma^\alpha_\beta \nabla \phi_\beta + \sum_{\beta,\gamma} w^\alpha_{\beta\gamma} \phi_\beta \nabla \phi_\gamma + ... + \bfeta_\alpha\Big],\label{eq:general-dynamics-1}
\end{align}
where ``..." represents higher-order terms and $\bfeta_\alpha$ is a Gaussian white noise with zero mean. In general, the noise may be multiplicative~\cite{kawasaki_stochastic_1994,dean_langevin_1996}, but, for simplicity, we work with additive noise such that
\begin{align}
    \langle \eta^i_\alpha(\br,t) \eta^j_\beta(\br',t')\rangle &= 2 D_\alpha \delta_{\alpha\beta} \delta_{ij} \delta^d(\br-\br') \delta(t-t')\;,
\end{align}
and defer discussions of multiplicative noises~\cite{noauthor_see_nodate}.
Equation~\eqref{eq:general-dynamics-1} describes systems whose departure from equilibrium stems either from the noise, as in multi-temperature systems, or from a non-conservative deterministic drift, as in non-reciprocal mixtures.
In the absence of nonlinearities the steady-state distribution is Gaussian, with short-range two-point correlations and vanishing higher cumulants. 

The role of nonlinearities can be explored by a simple scaling argument~\cite{tauber_critical_2014}: Under rescalings $\br \to b \br$, $t \to b^z t$, and $\phi_\alpha \to b^\zeta \phi_\alpha$, the linear terms in Eq.~\eqref{eq:general-dynamics-1} remain invariant with choice of $z=2$ and $\zeta=-d/2$. The leading nonlinear term now scales 
as $w^\alpha_{\beta\gamma} \to b^{-d/2} w^\alpha_{\beta\gamma}$ and is dimensionally irrelevant. 
Because all correlations are short-ranged in the linear theory, conventional reasoning~\cite{grinstein_conservation_1990} suggests that the irrelevant nonlinearities should not increase their range.

\begin{figure}
    \centering
    \includegraphics[]{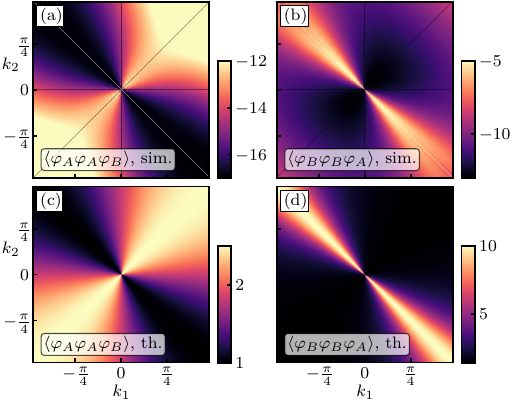}
    \caption{\textbf{Long-range 3-point correlations in a 2-temperature mixture.} The 3-point correlation from {\bf (a)-(b)} particle simulations and {\bf (c)-(d)} our theory [Eq.~\eqref{eq:2temp-3pt-equaltime}] at wavevectors $k_1$, $k_2$, and $k_3=-k_1-k_2$ in $d=1$. Panels~(a),~(c) represent species $\alpha=\beta=A,\,\gamma=B$, and panels~(b),~(d) represent species $\alpha=\beta=B,\,\gamma=A$. In the theory, we use parameters $\chi_3=-1$, $T_A=1$, and $T_B=0.1$. The simulations are the same as those shown in Fig.~\ref{fig:2temp-3pt-realspace}.}
    \label{fig:2temp-3pt}
\end{figure}

Here, we show that this reasoning fails. While symmetries enforce that two-point correlations remain short-ranged in these systems, higher-order correlations behave qualitatively differently. 
In particular, we find that generic nonequilibrium mixtures exhibit long-range three-point correlation functions decaying as $r^{-2d}$, corresponding to a discontinuity of their Fourier transforms in the low-wavenumber limit~\footnote{Following standard practice~\cite{schmittmann_driven_1998,torquato_local_2003,lefevere_high-temperature_2005,vinutha_stressstress_2023}, we designate correlations as ``long-range'' when they display a discontinuity or divergence in the low-wavenumber limit in momentum space. For $d$-dimensional systems, this corresponds to algebraic $n$-point correlation functions that decay at least as slowly as $r^{-(n-1)d}$. See~\cite{noauthor_see_nodate} for further discussion}.
These results are illustrated using numerical simulations of a microscopic model of two-temperature mixtures~\cite{awazu_segregation_2014,grosberg_nonequilibrium_2015,weber_binary_2016,smrek_small_2017,tanaka_hot_2017,ilker_phase_2020,wang_tethered_2021,jardat_diffusion_2022,schwarcz_emergence_2024,mccarthy_demixing_2024,damman_algebraic_2024,venkatareddy_growth_2025}, shown in Fig.~\ref{fig:2temp-3pt-realspace}(c). 
Long-range correlations are shown in real space (Fig.~\ref{fig:2temp-3pt-realspace}a,b) as well as in Fourier space (Fig.~\ref{fig:2temp-3pt}a,b). 
Our theory shows these long-range correlations to arise from nonlinearities that are formally irrelevant in the renormalization-group sense, but nonetheless control the large-scale behavior. 
All numerical details can be found in End Matter, and detailed computations are provided in~\cite{noauthor_see_nodate}.

\vspace{0.1in}
\noindent\textbf{\textit{Two-point correlations are short-ranged.}} 
First, we argue that symmetries forbid long-range 2-point correlations in Eq.~\eqref{eq:general-dynamics-1}, deferring a more detailed argument to~\cite{noauthor_see_nodate}. 
Throughout the Letter, we work primarily in Fourier space, where the structure of the theory is particularly transparent. We introduce the wavevector $\bk$ and study the dynamics of the field $\phi_\alpha(\bk,t)$. 
For an isotropic, translationally invariant system, the equal-time correlator $\langle \phi_\alpha(\bk,t) \phi_\beta(\bk',t)\rangle$ can only depend on $k=|\bk|$. 
It is given by the ratio of two functions of $k$: the numerator resulting from the noise correlation and the denominator resulting from the linear deterministic dynamics~\cite{schmittmann_statistical_1995}. 
Because the fields are conserved and diffusive, both the noise correlations and the deterministic dynamics contribute factors of $k^2$ at small wavenumbers. 
These cancel each other, resulting in a well-defined limit, $\sim k^0$, as $\bk\to 0$, which corresponds to short-range correlations. 

Note that this does not rule out the existence of short-range power-law correlations, as reported in Ref.~\cite{damman_algebraic_2024}. To find these, one must calculate the 2-point correlator at leading order in the nonlinearity, as detailed in~\cite{noauthor_see_nodate}. The leading-order correction scales as $k^d$, which results in a short-range two-point function decaying as $r^{-2d}$. Note that the threshold exponent distinguishing short- and long-range correlations in real space depends on the order of the correlation function.

While two-point functions suggest short-range correlations, we  show below that higher-order correlations are, however, long-ranged.
Three-point correlations play an important role in many contexts, ranging from driven-diffusive systems~\cite{hwang_three-point_1991,hwang_three-point_1993,lefevere_high-temperature_2005} to cosmology~\cite{takada_three-point_2003,sefusatti_cosmology_2006,desjacques_large-scale_2018}, turbulence~\cite{lii_bispectral_1976,manz_bispectral_2008,schmidt_bispectral_2020}, and quantum systems~\cite{daix_probing_2025,tam_singular_2026}. 
In isotropic, single-component nonequilibrium systems, three-point functions can exhibit short-ranged algebraic decay in $d\geq 2$~\cite{grinstein_generic_1991}, and are generally expected to be exponentially decaying in 1D. By contrast, we show that in multi-component nonequilibrium mixtures, three-point correlations become genuinely long-ranged, exhibiting a discontinuity in the low-wavenumber limit, even in 1D.

Before tackling the generic case described by Eq.~\eqref{eq:general-dynamics-1}, we study the simpler case of a phenomenological theory for multi-temperature mixtures, where the departure from equilibrium stems solely from the noises.

\vspace{0.1in}
\noindent\textbf{\textit{Multiple temperatures.}}
We consider two fields $\varphi_A$ and $\varphi_B$ connected to thermal baths of temperatures $T_A\neq T_B$. Introducing $\alpha,\beta,\gamma\in\{A,B\}$, we write their dynamics as
\begin{align}
    \dot{\phi}_\alpha &= \nabla \cdot \Big[ \nabla \Big(T_\alpha \phi_\alpha + \sum_\beta \chi_2 \phi_\beta + \sum_{\beta,\gamma} \chi_3 \phi_\beta \phi_\gamma\Big) + \bfeta_\alpha\Big]\;,\label{eq:2temp-phenom-dynamics}
\end{align}
where $\bfeta_\alpha(\br,t)$ is a Gaussian white noise field with covariance $\langle \eta^i_\alpha (\br,t) \eta^j_\beta(\br',t')\rangle = 2 T_\alpha \delta_{\alpha\beta} \delta_{ij} \delta^d(\br-\br') \delta(t-t').$ 
When $T_A=T_B$, the system is equivalent to equilibrium model B~\cite{tauber_critical_2014}. 
For simplicity, we have assumed that the virial coefficients $\chi_2$ and $\chi_3$ do not depend on the species of interacting fields $\alpha,\,\beta,$ and $\gamma$. Our results are robust to relaxing this assumption, and thus apply more broadly to two-temperature mixtures~\cite{grosberg_nonequilibrium_2015,ilker_phase_2020,damman_algebraic_2024}.

Because the nonlinearities are irrelevant in this theory by scaling, one might expect the correlations to be short-range as in the linear model. To show that this is not the case, we evaluate the three-point correlation in the limit of weak interactions, treating the nonlinearity $\chi_3\sim {\cal O}(\epsilon)$ perturbatively. For simplicity, we also take $\chi_2$ to scale like $\epsilon$, although this doesn't change our essential result. To leading order, this yields the 3-point correlation~\cite{noauthor_see_nodate}
\begin{align}
    \langle &\phi_\alpha (\bk_1) \phi_\beta(\bk_2) \phi_\gamma(\bk_3)\rangle \label{eq:2temp-3pt-equaltime}\\
    &= - 2\chi_3 \frac{(2\pi)^{d} \delta^d(\bk_1+\bk_2+\bk_3)(k_1^2 + k_2^2 + k_3^2)}{T_\alpha k_1^2 + T_\beta k_2^2 + T_\gamma k_3^2} + \mathcal{O}(\epsilon^2)\;.\nonumber
\end{align}
When, for example, $\alpha\neq \beta$ and $T_\alpha\neq T_\beta$, this expression is scale-free and discontinuous in the limit $k_i \to 0$, indicating long-range correlations consistent with an $r^{-2d}$ scaling. By contrast, when $T_\alpha=T_\beta=T_\gamma$, the expression becomes regular, consistent with short-range behavior in equilibrium. 
Equation~\eqref{eq:2temp-3pt-equaltime} is plotted in Fig.~\ref{fig:2temp-3pt}c,d as a function of the wavevectors $k_1$ and $k_2$ (with $k_3 = -k_1 - k_2$). Its ``pinch-point" structure resembles the 2-point correlations of conserved anisotropic systems~\cite{schmittmann_statistical_1995,schmittmann_driven_1998}, which also have two independent wavevectors $(k_\parallel,k_\perp)$. Similarly, lattice anisotropy leads to a similar pinch-point structure in the Coulomb phase of frustrated lattices with conservation laws~\cite{henley_coulomb_2010}. However, the phenomenon illustrated by Eq.~\eqref{eq:2temp-3pt-equaltime} is fundamentally different from anisotropy-induced long-range correlations, since it does not affect the 2-point correlator and is absent in theories with a single field. Moreover, it appears even in 1 spatial dimension, and does not require the imposition of spatial anisotropy; rather, it emerges generically in nonequilibrium systems with multiple species.

To examine the real-space structure of these correlations, we Fourier-transform Eq.~\eqref{eq:2temp-3pt-equaltime}. As long as the temperatures are not all equal~\cite{noauthor_see_nodate}, we find
\begin{align}
    \langle& \phi_\alpha(\br_1) \phi_\beta(\br_2) \phi_\gamma(\br_3)\rangle \\
    &\propto (\nabla_1^2 + \nabla_2^2 + \nabla_3^2) \big[T_\gamma r_{12}^2 + T_\alpha r_{23}^2 + T_\beta r_{31}^2\big]^{1-d}+\mathcal{O}(\epsilon^2)\,,\nonumber
\end{align}
where $r_{ij} \equiv |\br_i-\br_j|$. When $d=1$, the $[\ldots]^{1-d}$ is replaced by $\log[\ldots]$. The correlator decays like $r^{-2d}$ in space.  Physically, this behavior reflects the emergence of long-range, quadrupolar density modulations: two localized fluctuations of the fields $\phi_\alpha$ and $\phi_\beta$ induce a slowly decaying response in the density of the third field $\phi_\gamma$.

\vspace{0.1in}
\noindent\textbf{\textit{Comparison between phenomenological theory and numerical simulation.}} To test the predictions of the phenomenological field theory given by Eq.~\eqref{eq:2temp-phenom-dynamics}, we simulate a binary mixture of overdamped Brownian particles in $d=1$, with particle $i$ of species $\alpha$ having temperature $T_i=T_\alpha$. This system is illustrated in Fig.~\ref{fig:2temp-3pt-realspace}c. The particles interact through the soft repulsive potential $U_{\rm int}(x) = \Theta(1-|x|) \frac{\kappa}{2} \big(1 - |x|\big)^2$ with dynamics
\begin{align}
    \dot{x}_i(t) &= -\sum_{j\neq i} U_{\rm int}'(x_i-x_j) + \sqrt{2 T_i} \eta_i(t)\,,
\end{align}
where $\eta_i(t)$ is a zero-mean Gaussian noise with variance $\langle \eta_i(t) \eta_j(t')\rangle = \delta_{ij} \delta(t-t')$. 

We observed the predicted long-ranged decay of three-point correlations in real space, $\sim r^{-2}$, as shown in Fig.~\ref{fig:2temp-3pt-realspace}a,b. The three-point correlations in Fourier space are displayed in Fig.~\ref{fig:2temp-3pt}a,b. Up to an overall shift, they bear strong qualitative resemblance to the theory shown in Fig.~\ref{fig:2temp-3pt}c,d, corresponding to Eq.~\eqref{eq:2temp-3pt-equaltime}, despite the purely phenomenological nature of Eq.~\eqref{eq:2temp-phenom-dynamics}. Details of the non-analytic structure at small wave vectors can be found in~\cite{noauthor_see_nodate}, together with contrasting data for the single-temperature case and the numerical demonstration that 2-point correlations are short-ranged.

\vspace{0.1in}
\noindent\textbf{\textit{General theory.}}
The emergence of long-range three-point correlations is not specific to the two-temperature model, but holds for the general dynamics of  Eq.~\eqref{eq:general-dynamics-1}. 
To see this, it is convenient to diagonalize the linear part of the dynamics. This yields independent diffusive modes $\psi_\alpha$ coupled through nonlinear interactions and correlated noise, evolving as~\cite{noauthor_see_nodate}
\begin{align}
    \dot{\psi}_\alpha &= \nabla \cdot \Big[m_\alpha \nabla \psi_\alpha + \sum_{\beta,\gamma} u^\alpha_{\beta\gamma} \psi_\beta \nabla \psi_\gamma + \bLambda_\alpha\Big]\;,\label{eq:general-dynamics-diag}
\end{align}
with $m_\alpha$, $u^\alpha_{\beta\gamma}$, and $D_{\alpha\beta}$ defined as transformed parameters, and where $\bLambda_\alpha(\br,t)$ is a Gaussian white noise field whose covariance reads
\begin{align}
    \langle \Lambda^i_\alpha &(\br,t) \Lambda^j_\beta(\br',t')\rangle = 2D_{\alpha\beta} \delta_{ij} \delta(t-t') \delta^d(\br-\br')\;.\label{eq:Lambda}
\end{align}

Perturbative calculation of the three-point correlations, detailed in~\cite{noauthor_see_nodate}, proceeds as before and, to lowest order in the couplings $u^\alpha_{\beta\gamma} \sim \epsilon$~\footnote{Treating $u^\alpha_{\beta\gamma}$ perturbatively is justified since it is an irrelevant operator.}, we find 
\begin{widetext}
    \begin{align}
        \langle \psi_\alpha &(\bk_1) \psi_\beta(\bk_2) \psi_\gamma(\bk_3) \rangle = \frac{4 (2\pi)^d \delta^d(\bk_1+\bk_2+\bk_3)}{m_\alpha k_1^2 + m_\beta k_2^2 + m_\gamma k_3^2} \label{eq:lr-general-ans-fourier}\\
        &\times \sum_{\mu,\lambda} \Bigg[\frac{D_{\beta\mu} D_{\gamma\lambda} \bk_1 \cdot \big[u^\alpha_{\mu\lambda} \bk_3 + u^\alpha_{\lambda\mu} \bk_2\big]}{(m_\beta+m_\mu)(m_\gamma+m_\lambda)} + \frac{D_{\alpha\mu} D_{\gamma\lambda} \bk_2 \cdot \big[u^\beta_{\mu\lambda} \bk_3 + u^\beta_{\lambda\mu} \bk_1\big]}{(m_\alpha+m_\mu)(m_\gamma+m_\lambda)} + \frac{D_{\beta\mu} D_{\alpha\lambda} \bk_3 \cdot \big[u^\gamma_{\mu\lambda} \bk_1 + u^\gamma_{\lambda\mu} \bk_2\big]}{(m_\beta+m_\mu)(m_\alpha+m_\lambda)}\Bigg] + \mathcal{O}(\epsilon^2)\;.\nonumber
    \end{align}
\end{widetext}
This generalization of Eq.~\eqref{eq:2temp-3pt-equaltime} indicates generic long-range correlations, and corresponding  $\sim r^{-2d}$ decay in real space. This exponent is unchanged at higher orders in $\epsilon$~\cite{noauthor_see_nodate}.

Equation~\eqref{eq:lr-general-ans-fourier} also allows us to establish the conditions under which the fields $\psi_\alpha$ and $\phi_\alpha$, related through local linear maps, exhibit short-range correlations. We find that this is the case whenever the dynamics correspond to equilibrium model B, i.e.~\cite{noauthor_see_nodate}
\begin{align}   \text{Equilibrium} \;\Leftrightarrow\;    
\begin{matrix}       
&w^\alpha_{\beta\gamma} = w^\alpha_{\gamma\beta} \\      
&\Gamma^\alpha_\beta{/D_\alpha} = \Gamma^\beta_\alpha{/D_\beta}\\      
&w^\alpha_{\gamma\beta}{/D_\alpha} = w^\beta_{\gamma\alpha}{/D_\beta}
\end{matrix}
\quad \forall\,\alpha,\beta,\gamma\;. 
\end{align}

Note also that long-range correlations only appear when the fields are not all of the same species. Clearly, if we substitute $\alpha=\beta=\gamma$ into Eq.~\eqref{eq:lr-general-ans-fourier}, we find that the three-point function is short-range. However, it can also be argued on more general grounds: Because $\langle \psi_\alpha(\bk_1) \psi_\alpha(\bk_2) \psi_\alpha(\bk_3)\rangle$ must be symmetric with respect to permutations of $(\bk_1,\bk_2,\bk_3)$, to quadratic order, its numerator and denominator must be linear combinations of $k_1^2+k_2^2+k_3^2$ and $\bk_1\cdot\bk_2 + \bk_2\cdot\bk_3 + \bk_3\cdot\bk_1$. But by translational symmetry, $\bk_1+\bk_2+\bk_3=\mathbf{0}$; squaring this leads to $k_1^2+k_2^2+k_3^2=-2(\bk_1\cdot\bk_2 + \bk_2\cdot\bk_3 + \bk_3\cdot\bk_1).$ The leading-order $\bk_i$-dependence thus cancels, resulting in short-range correlations. 

Finally, note that in nonequilibrium mixtures with an up-down symmetry that prevents $\phi^3$ interactions, irrelevant $\phi^4$ interactions induce long-range 4-point correlations with an exponent $r^{-3d}$~\cite{noauthor_see_nodate}. Presumably, $\phi^n$ interactions induce long-range $n$-point correlations for all $n>2$.

\vspace{0.1in}
\noindent\textbf{\textit{Conclusion.}}
Thirty-five years after the seminal work of Grinstein, Lee, and Sachdev on long-ranged correlations in anisotropic nonequilibrium systems~\cite{grinstein_conservation_1990}, we have identified a new mechanism that induces long-ranged correlations out of equilibrium. Namely, we have shown how dangerously irrelevant operators can lead to true long-ranged correlations in the three-point correlation functions of nonequilibrium mixtures. 
Our derivation only relies on a perturbative treatment of nonlinear terms, which is justified due to their irrelevant nature. 
It also explains the emergence of short -ranged algebraic decay in two-point functions. 
Note that multiplicative noises enter at the same order as the nonlinear terms. For simplicity, we considered additive noises in this Letter, which also justifies the choice of constant mobility. In~\cite{noauthor_see_nodate}, we show how nonequilibrium multiplicative noises can also induce long-ranged three-point functions, with the same exponent, showing the robustness of our results. The ordered and pattern-forming phases of active mixtures have rightfully attracted a lot of attention~\cite{you_nonreciprocity_2020,saha_scalar_2020,fruchart_non-reciprocal_2021,dinelli_non-reciprocity_2023,pisegna_emergent_2024,fruchart_nonreciprocal_2026}; our results highlight that their disordered phases also display a surprisingly rich phenomenology.

Understanding the physics of active mixtures is essential for understanding the complex physics of living systems, in particular, the cell~\cite{lingwood_lipid_2010,gowrishankar_active_2012,ganai_chromosome_2014,zidovska_rich_2020,goychuk_polymer_2023,rautu_active_2026}. While significant effort has been devoted to understanding the formation of biomolecular condensates~\cite{jacobs_phase_2017,weber_physics_2019}, genomic spatial organization~\cite{ganai_chromosome_2014,goychuk_polymer_2023}, and other ordered phases, a significant portion of the cell is occupied by a disordered mixture of proteins, enzymes, and other biomolecules, driven out of equilibrium by chemical imbalances and active processes. 
Our results reveal new mechanisms for long-range correlations within this disordered bath that may help increase our understanding of the complex web of information transfer in living systems.

\vspace{0.1in}
\noindent\textbf{\textit{Acknowledgments}}:
We thank Vincent D\'emery and J\'er\'emy O'Byrne for useful discussions, as well as an anonymous referee for drawing our attention to the case of multiplicative noises. YK acknowledges financial support from ISF (2038/21), (3457/25). YK and MK acknowledge financial support from NSF/BSF (DMR-2022605). This research was supported in part by grant NSF PHY-2309135 to the Kavli Institute for Theoretical Physics (KITP).

\bibliography{refs}

\newpage 

\vspace{0.1in}
\noindent\textbf{\textit{End Matter.}}
For the particle simulations depicted in both figures (Figs.~\ref{fig:2temp-3pt-realspace}-\ref{fig:2temp-3pt}), we simulate $N_A=200$ Brownian particles of temperature $T_A=1$ and $N_B=200$ Brownian particles of temperature $T_B=0.1$ in a 1-dimensional system of size $L=800$. The position $x_i(t)$ of particle $i$ is numerically integrated using an Euler scheme with timestep $dt=5 \times 10^{-4}$, given by
\begin{widetext}
\begin{align}
    1 \leq i \leq N_A:\qquad x_i(t+dt) &= x_i(t) - dt\sum_{j \neq i} U_{\rm int}'\big(x_i(t)-x_j(t)\big) + \sqrt{2 T_A dt} \,\eta_i^t\\
    N_A < i \leq N_A+N_B:\qquad x_i(t+dt) &= x_i(t) - dt\sum_{j \neq i} U_{\rm int}'\big(x_i(t)-x_j(t)\big) + \sqrt{2 T_B dt} \,\eta_i^t
\end{align}
\end{widetext}
where $\eta_i^t$ is a Gaussian random variable with zero mean and correlations $\langle \eta_i^t \eta_j^s\rangle = \delta_{i,j} \delta_{t,s}$. The particles interact through the soft repulsive harmonic potential
\begin{align}
    U_{\rm int}(x) &= \Theta(1-|x|) \frac{\kappa}{2} \big(1 - |x|\big)^2\;,
\end{align}
where we use $\kappa=1$ in all simulations.

We start the system in a random uniform configuration and 
measure the 3-point correlations
\begin{align}
\langle\phi_A(k_1)\phi_A(k_2)\phi_B(-k_1-k_2)\rangle\,\quad\text{and}\notag\\
\langle \phi_B(k_1)\phi_B(k_2)\phi_A(-k_1-k_2)\rangle\;,\end{align}
for wavenumbers $k_1=2\pi n_1/L$ and $k_2=2\pi n_2/L$ with $n_1,n_2 \in \{-600,-599,\ldots,600\}$, every $\Delta t=1$. The statistics shown in this Letter were obtained using a final time $t_{\rm f}={2\times} 10^6$. Because the system is disordered, equilibration at the length-scales $\ell \sim 100$ displayed in Figs.~\ref{fig:2temp-3pt-realspace}-\ref{fig:2temp-3pt} occurs on a timescale $\tau_{\rm eq} \sim \ell^2/(2 \frac{T_A+T_B}{2}) \sim 100^2 / 2 \sim 10^4 \ll t_{\rm f}$; thus, we average data collected over the entire simulation period without a separate equilibration period. We repeat this for 840 independent realizations, and average the 3-point correlations over these different realizations. 

To calculate the real-space 3-point correlation (Fig.~\ref{fig:2temp-3pt-realspace}a,b), we perform an inverse Fourier transform of the 3-point correlation measured in Fourier-space. Because the Fourier-space correlation is real and symmetric under $(k_1,k_2) \to (-k_1,-k_2)$, this is given by
\begin{widetext}
\begin{align}
    \langle \phi_\alpha(x_1) \phi_\beta(x_2) \phi_\gamma(x_3)\rangle &= \langle \phi_\alpha(x_1-x_3) \phi_\beta(x_2-x_3) \phi_\gamma(0)\rangle\notag \\
    &= \frac{1}{L^2} \sum_{k_1,k_2} \cos \big[k_1 (x_1-x_3) + k_2 (x_2-x_3)\big] \langle \phi_\alpha(k_1) \phi_\beta(k_2) \phi_\gamma(-k_1-k_2)\rangle\;.
\end{align}
\end{widetext}
This forms a grid in the $(x_{13},x_{23})=(x_1-x_3,x_2-x_3)$ space. In Fig.~\ref{fig:2temp-3pt-realspace}a,b, we plot slices from this grid defined by $x_{13} = n x_{23}$ for $n=2,3,4$.

\nocite{hwang_three-point_1991,korniss_long_1997,schmittmann_driven_1998,torquato_local_2003,lefevere_high-temperature_2005,torquato_structural_2021,adachi_activity-induced_2022,vinutha_stressstress_2023,ramaswamy_active_2003,dey_spatial_2012,lighthill_introduction_1958,bleistein_asymptotic_1986,tauber_critical_2014,damman_algebraic_2024,symanzik_small_1970,callan_broken_1970,symanzik_small-distance-behaviour_1971,obyrne_nonequilibrium_2023,dinelli_non-reciprocity_2023,obyrne_geometric_2025}

\end{document}